\documentclass[aps,prb,twocolumn,showpacs,floatfix,superscriptaddress,nofootinbib]{revtex4-2}
\usepackage{graphicx}%
\usepackage{dcolumn}%
\usepackage{bm}%
\usepackage{physics}
\usepackage{color}
\usepackage{amsmath}
\usepackage{amssymb}
\usepackage{ulem} %
\usepackage{multirow}
\usepackage{booktabs}
\usepackage{blkarray}

\definecolor{gr4} {RGB}{34,118,34}

\newcommand{\Snum}{S_{\text{num},d}}
\newcommand{\Sconf}{S_{\text{conf},d}}

\newcommand{\xiloc}{\xi_{\text{loc}}}

\begin{document}
\title{Tracking locality in time evolution of disordered systems}%
\author{Tomasz Szo\l{}dra} 
\affiliation{Instytut Fizyki Teoretycznej, 
Uniwersytet Jagiello\'nski,  \L{}ojasiewicza 11, PL-30-348 Krak\'ow, Poland}
\author{Piotr Sierant} 
\affiliation{ICFO-Institut de Ci\`encies Fot\`oniques, The Barcelona Institute of Science and Technology, Av. Carl Friedrich
Gauss 3, 08860 Castelldefels (Barcelona), Spain}
\author{Maciej Lewenstein} 
\affiliation{ICFO-Institut de Ci\`encies Fot\`oniques, The Barcelona Institute of Science and Technology, Av. Carl Friedrich
Gauss 3, 08860 Castelldefels (Barcelona), Spain}
\affiliation{ICREA, Passeig Lluis Companys 23, 08010 Barcelona, Spain}
\author{Jakub Zakrzewski} 
\affiliation{Instytut Fizyki Teoretycznej, 
Uniwersytet Jagiello\'nski,  \L{}ojasiewicza 11, PL-30-348 Krak\'ow, Poland}
\affiliation{Mark Kac Complex Systems Research Center, Uniwersytet Jagiello{\'n}ski, PL-30-348 Krak{\'o}w, Poland}

\date{\today}

\begin{abstract}
Using local density correlation functions for a one-dimensional spin system, we introduce a correlation function difference (CFD) which compares correlations on a given site between a full system of size $L$ and its restriction to $\ell<L$ sites. We show that CFD provides useful information on transfer of information in quantum many-body systems by considering the examples of ergodic, Anderson, and many-body localized regimes in disordered XXZ spin chain. In the ergodic phase, we find that the propagation of CFD is asymptotically faster than the spin transport but slower than the ballistic propagation implied by the Lieb-Robinson bound. In contrast, in the localized cases, we unravel an exponentially slow relaxation of CFD. Connections between CFD and other observables detecting non-local correlations in the system are discussed.
\end{abstract}
\maketitle

\section{Introduction}

The understanding of dynamics of interacting quantum many-body systems has been a subject of extensive research over the recent years. The eigenstate thermalization hypothesis \cite{Deutsch91,Srednicki94,Rigol08, Alessio16} anticipates that such systems are spectrally faithful to random matrix theory predictions, and when prepared in out-of-equilibrium states, they evolve in time to an equilibrium state dependent only on few global constants of motion but independent of the details of the initial state. 
One exception to this ergodic paradigm is provided by a class of integrable models \cite{Vidmar16}. 
Similarly, ergodicity may be broken in the presence of a sufficiently strong disorder due to the phenomenon of many-body localization (MBL) \cite{Gornyi05,Basko06, Pal10}. MBL is characterized by a simultaneous occurrence of various features such as the emergent integrability of the system (with the existence of a complete set of local integrals of motion \cite{Huse14,Ros15}) resulting in Poissonian spectral statistics \cite{Serbyn16, Buijsman18, Sierant19b, Sierant20, Rao21, De21, Rao22}, area-law entangled eigenstates \cite{Serbyn13b, Bauer13}, logarithmic growth of the entanglement entropy in time dynamics of initially low-entangled states \cite{Znidaric08, Bardarson12, serbyn2013universal}, and a long memory of the initial state in the dynamics \cite{Nandkishore15,Bera17, Sierant18, Herbrych22} (for recent reviews see \cite{Alet18,Abanin19}). Interestingly, last years have brought new developments to our understanding of non-equilibrium quantum many-body physics. On one side, nonergodic dynamics has been observed in disorder-free systems such as tilted lattices \cite{Schulz19, vanNieuwenburg19, Taylor20, Chanda20c, Yao20b, Scherg21, Guo21, Morong21, Yao21, Yao21a}, in models with global constraints \cite{Turner18,Feldmeier19,Nandkishore19,Sala20,Khemani20,Sierant21,Szoldra22}, or, notably, in implementations of lattice gauge theories \cite{Robinson19,James19,Surace19,Chanda20,Banuls20,Aidelsburger22,Aramthottil22}. On the other side, following a seminal work \cite{Suntajs20e}, an intensive debate developed concerning the very existence of MBL in the thermodynamic limit \cite{Panda20,Abanin21,Sierant20b,Sierant20p,Morningstar22,Crowley22,Long22}. 
While it seems that there exist models showing a genuine transition to MBL phase that survives the thermodynamic limit \cite{Sierant22f,Suntajs22s}, the paradigmatic model for MBL, i.e.  the disordered Heisenberg  chain, still defies a definite answer. Some recent works  \cite{Morningstar22, Sels22} suggest that a stable MBL phase occurs in this model only at a  very strong disorder shifting the previously considered border between ergodic and MBL phase \cite{Luitz15, Sierant20p} by an almost one order of magnitude in disorder amplitude. 
Other studies negate altogether the existence of MBL highlighting  a presence of a persistent (but extremely slow) growth of the classical part of the entanglement entropy, the so called number entropy \cite{Kiefer20,Kiefer21}  (see, however \cite{Ghosh22} for an opposite interpretations of those results).

\begin{figure}
    \centering
    \includegraphics[width=.75\columnwidth]{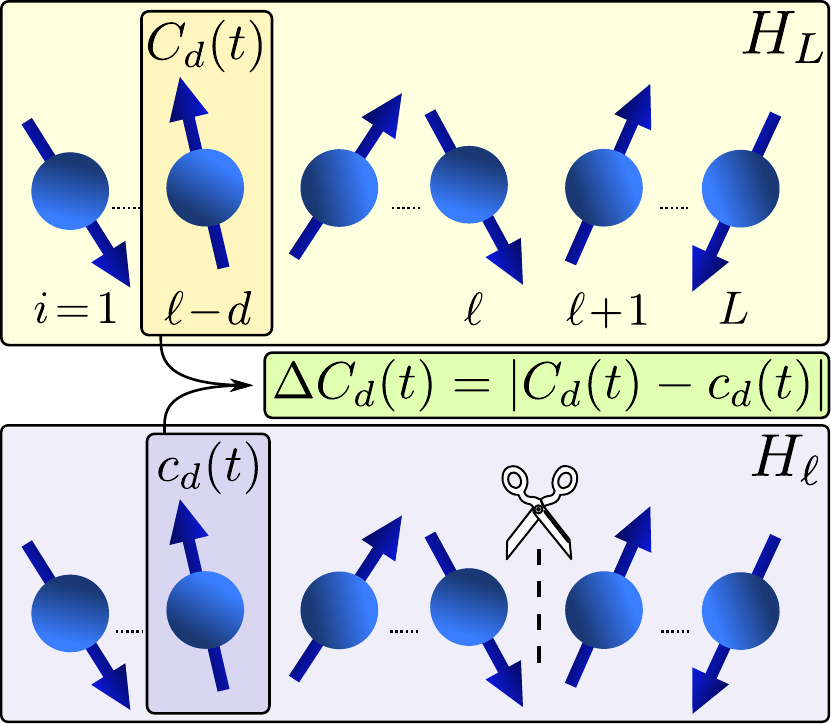}
    \caption{Scheme for tracking locality of quantum dynamics in a spin lattice. We compare the time evolution of the single-site 
    correlation function $C_d(t)$ at site $i= l-d$ in a system comprised out of $L$ sites (upper panel) with $c_d(t)$ at the same site but in a system of $\ell <L$ sites (bottom panel). 
    }
    \label{fig:scheme}
\end{figure}
While often sophisticated measures (like an entanglement entropy or 
a spectral form factor) are considered, it seems worthwhile to focus also on observables more easily accessible experimentally, such as the imbalance addressed in the first experiments showing signatures of MBL \cite{Schreiber15,Luschen17} or the corresponding two-point density correlation functions.
Still, even for those simpler observables, it can be shown that no conclusive answer about the stability of MBL phase can be obtained on timescales offered by present day experiments as well as state-of-the-art simulations using tensor networks \cite{Sierant22c}.

The aim of this work is to introduce an observable that allows one to probe the locality of dynamics in quantum many-body systems that would be experimentally more accessible than the out-of-time-order correlator (OTOC) function, conventionally employed in studies of thermalization and quantum information scrambling \cite{Sekino08, Maldacena16, Maldacena16a,Luitz17inf, Bohrdt17, Luitz19} 
(see \cite{Huang17, Fan17, Swingle17, He17, Smith19} for studies of disordered systems) or spin echo and double electron-electron resonances (DEER) employed in \cite{Serbyn14a}. To that end, we consider a 1D lattice system comprised of $L$ sites with open boundary conditions (OBC) and we examine two settings: the first, in which we calculate the two times density correlation function $C_d(t)$ in a given site (later called simply as density correlation function for the full system of $L$ sites; the second in which we compute the similar density correlation function $c_d(t)$ for a system restricted to $l<L$ sites. Our setup is illustrated in Fig.~\ref{fig:scheme}. The observable of our interest is a correlation function difference (CFD), $\Delta C_d(t)$, calculated as
\begin{equation}
    \Delta C_d(t) = \left| C_d(t) - c_d(t)\right|,
    \label{eq:cfd}
\end{equation} 
see Sec.~\ref{sec:model} for a precise definition of $C_d(t)$, $c_d(t)$.
The CFD reflects changes in the time evolution of density correlation functions that occur when the system size is increased from $\ell$ to $L$. The realization of random disorder on the first $\ell$ sites stays the same. This quantifies the locality of quantum dynamics: if the dynamics at site $d$ is not affected by degrees of freedom residing on sites separated from $d$ by a certain length scale $\xi$, we expect a vanishing CFD $\Delta C_d(t)\approx 0$ as long as $l-d > \xi$. In contrast, a non-zero CFD indicates a non-trivial impact of the degrees of freedom at sites $i> \ell$ on time evolution at site $d$.

We note that CFD is particularly relevant for experiments in which $C_d(t)$ is directly measurable \cite{Rispoli19,Lukin19} as it requires a density measurement with a single site resolution. Moreover, the change of the system size from $\ell$ to $L$ in Fig.~\ref{fig:scheme} can be alternatively interpreted as an introduction of a local tunneling term that couples sites $i<\ell$ and $i \geq \ell$ to the Hamiltonian for $L$ sites. This provides a conceptual link between CFD and earlier studies of deformation of many-body eigenstates due to local perturbations \cite{Sierant19, Pandey20, Maksymov19, Garratt22}.
A similar scenario was considered also in the context of local quench of a one-dimensional quantum spin chain in which two parts of the chain are joined \cite{Calabrese06, Eisler07, Stephan11} or split \cite{Zamora14}. In that case, the dynamics and the relevant time scales are dictated by the physics of low energy excitations (in particular, by the underlying conformal field theory for critical spin chains). In contrast, in this work we probe dynamics of highly out-of-equilibrium initial states.

Section~\ref{sec:model} contains the definition of disordered Heisenberg model on which we benchmark CFD. In Section~\ref{sec_res} we study the dynamics of the model for three distinct regimes: ergodic phase with delocalized eigenstates, Anderson localized (non-interacting) model and MBL regime.  We show that CFD behaves differently in each of the three regimes, reflecting the various degrees of locality of dynamics in the system. We also relate our observable to the number entropy which measures the fluctuations of the number of particles in different parts of the system. Finally, we conclude in Section~\ref{sec_con}.

\section{The model}
\label{sec:model}
To study characteristic properties of CFD in different dynamical phases of matter, we consider the disordered XXZ spin chain, with the Hamiltonian:
\begin{align}
    H_{L} &= \sum_{i=1}^{L-1} J\left(S^x_{i} S^x_{i+1} + S^y_{i} S^y_{i+1}\right) + \Delta S^z_{i} S^z_{i+1} \nonumber \label{eq:xxz} \\&\qquad\qquad+ \sum_{i=1}^L h_i S^z_i, 
\end{align}
where $S^k_i$, $k=x,y,z$ are spin-1/2 operators at site $i$ and the on-site random magnetic field $h_i$ is uniformly distributed in {the} $[-W, W]$ interval{ where $W$ is the disorder amplitude}. Further on, we take $J=1$ as an energy unit. The model \eqref{eq:xxz} maps to spinless interacting fermions via Jordan-Wigner transformation where $J$ corresponds to tunneling amplitude between sites while $\Delta$ gives the strength of nearest-neighbor interactions. 

We shall employ CFD to investigate the process of quantum information transfer in this system and compare the three regimes: 
ergodic phase ($\Delta=1$, $W<W_c$), Anderson-localized phase ($\Delta=0$, $W>0$), and the MBL regime ($\Delta=1$, $W>W_c$), where $W_c$ is the critical disorder strength at which, at relatively small system sizes, the crossover between delocalized and MBL regimes occurs in an interacting system. At the system sizes considered, $L\approx 20$, the crossover is located around $W_c\approx 3.5$ \cite{Luitz15, Sierant20p}. 
We study time dynamics from an initial N\'eel state, a separable state with every second spin up and every second down (we limit the analysis to a conserved total magnetization sector $S_z=0$ for even $L$) corresponding to every second site occupied and every second site empty in the fermionic language.
While the early experiments \cite{Schreiber15} considered a global observable, i.e. a staggered magnetization
(in other words the imbalance between odd and even sites) over a whole lattice, we consider
the single site correlation functions defined as $C_i(t) = \overline {C^{\mathrm{s}}_i(t) }$, where the overline denotes the disorder average and $C^{\mathrm{s}}_i(t)=\expval{S^z_i(t)S^z_i(0)}$ is the quantum mechanical expectation value for the $i$-th spin at time $t$.
The standard imbalance is proportional to a sum of all $C_i(t)$. Here, we use the single site correlation functions to 
calculate the CFD according to Eq.~\eqref{eq:cfd}.
To evaluate $c_i(t)$, we modify the Hamilotnian, \eqref{eq:xxz}, by removing the link between sites $i=\ell, \ell+1$, i.e. by removing the term $J\left(S^x_{\ell} S^x_{\ell+1} + S^y_{\ell} S^y_{\ell+1}\right) + \Delta S^z_{\ell} S^z_{\ell+1}$.
Such a cut may be realized (probably with some experimental difficulty) using, for example, the technique of sub-wavelength optical barriers \cite{Jendrzejewski16}.
Alternatively, projection of site-resolved optical potentials \cite{Bakr09} in conjunction with optical boxes potentials \cite{Navon21} can be used to separately investigate dynamics of the system comprised of $l$ and $L$ sites. Similarly, CFD may be studied experimentally in other quantum simulators such as ion chains or Rydberg atom arrays - for a review see \cite{Altman21}.

\section{CFD in various dynamical regimes}
\label{sec_res}

\subsection{Ergodic phase}
\label{sec:erg}

%%%%%%%%%%%%%%%%%%%%%%%%%%%%%%%%%%%%%%%%%
%%%   fig. 1
%%%%%%%%%%%%%%%%%%%%%%%%%%%%%%%%%%%%%%%%%

\begin{figure}
    \centering
    \includegraphics[width=.85\columnwidth]{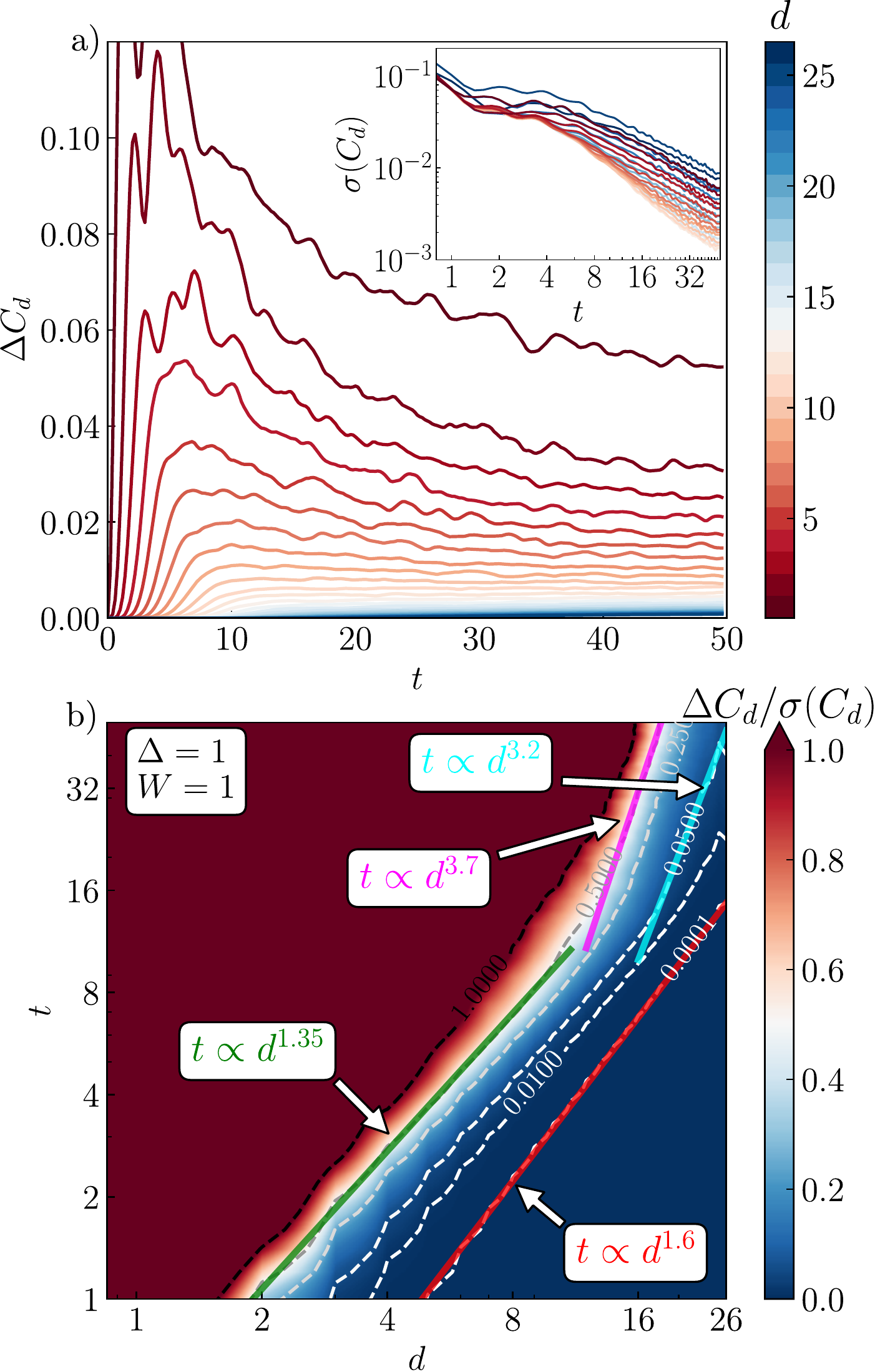}
    \caption{The CFD, $\Delta C_d$, in the ergodic phase of the XXZ chain with disorder strength $W=1$, length $L=28$ and a cut at $\ell=26$. Panel a) shows time dependence of $\Delta C_d$ for various distances $d$ from the cut. The inset presents the rolling standard deviation $\sigma(C_d)${, Eq.\eqref{eq:roll},} that quantifies temporal fluctuations of the single site correlation function $C_d(t)$. Panel b) shows the rescaled CFD, $\Delta C_d/\sigma(C_d)$ as function of the  distance from the cut and time $t$ {in the log-log scale}, the {white} dashed lines denote isovalues of $\Delta C_d/\sigma(C_d)$, numbers indicate the values for different lines. Those almost straight lines in the log-log plot are thus well fitted piecewise by a power law $t\propto d^\alpha$ with values of the exponent indicated in the figure. Averages are taken over $1000$ disorder realizations.
    }
    \label{fig:ergodic_lightcone}
\end{figure}

We start our investigation of CFD by considering the ergodic phase of the disordered XXZ spin chain, setting $W=1$ and $\Delta = 1$ in \eqref{eq:xxz}. We note that, in the fermionic interpretation of the model, the particle transport in the ergodic phase of the XXZ spin chain is, depending on the value of disorder strength $W$, diffusive for $0<W \lesssim 0.55$ (for $\Delta=1$) or sub-diffusive for larger disorder strengths $W<W_c$ \cite{BarLev15, Znidaric16, Bertini21}. Moreover, 
for a short-range Hamiltonian we study,  one expects that the velocity of information spreading in the system is bounded by the Lieb-Robinson velocity \cite{Lieb72, Baldwin22} which implies that local excitations propagate within a causal light cone.

With those two points as guiding principles, we compute the CFD, $\Delta C_d$, for XXZ spin chain at $W=1$ and $\Delta = 1$ by calculating the single site correlation functions $c_i(t)$ and $C_i(t)$ for system sizes $l=26$ and $L=28$, respectively, using the Chebyshev expansion of the time evolution operator \cite{Fehske08}. The resulting CFD, shown in Fig.~\ref{fig:ergodic_lightcone}a), becomes non-zero at times that increase with the distance $d$ from the cut. This is in accordance with the hypothesis that CFD probes correlations in the system that become increasingly non-local during the course of the time evolution. Importantly, the initial increase of  $\Delta C_d$ is followed by a damping of CFD at later times. The latter behavior does not reflect changes in the non-local content of the dynamics of the system, but rather is caused by the decrease of temporal fluctuations of $C_d(t)$ and $c_d(t)$ due to presence of interactions \cite{Serbyn14, Nandy21}.

In order to isolate this phenomenon, we quantify the temporal fluctuations of $C_d(t)$ ($c_d(t)$) by calculating a rolling standard deviation 
\begin{equation}
    \sigma(C_d)= \overline{ \left \langle (  C^{\mathrm{s}}_i(t) - 
   \left \langle C^{\mathrm{s}}_i(t) \right \rangle_{\Delta t} )^2  \right \rangle_{\Delta t}},
    \label{eq:roll}
\end{equation}
where the overline denotes the disorder average, and $\left \langle.\right \rangle_{\Delta t}$ is an average over time window of some small size $\Delta t$ around time $t$ (we take $\Delta t=5$, as checked the results are similar for $\Delta t=1,10$). The inset in Fig.~\ref{fig:ergodic_lightcone}a) shows that the decay of $\sigma(C_d)$ is approximately described by a power law in the interval of times considered. This decay of the amplitude of temporal oscillations of the single site correlation functions is the main factor responsible for the late time decrease of CFD. We shall discuss in details the mechanism of this effect in the following section. To compensate for this effect we inspect the dimensionless quantity $\Delta C_d/\sigma(C_d(t))$, which we refer to as the rescaled CFD. 

Figure~\ref{fig:ergodic_lightcone}b) shows the spatiotemporal dependence of the rescaled CFD in ergodic phase of the disordered XXZ spin chain. In particular, by considering the curves at which $\Delta C_d/\sigma(C_d(t)) = \mathrm{const}$ we find that a given value of the rescaled CFD is reached after a time scaling as a power law, $t \propto d^{\alpha}$, with the distance from the cut and $1<\alpha <2$. The power $\alpha$ (dependent on the isovalue curve as denoted in the figure) is consistently greater than unity for small and medium $d$ indicating that the information about the cut revealed by the rescaled CFD spreads superdiffusively. Thus, CFD is not related to a particle (spin) transport which is subdiffusive for the parameters used \cite{BarLev15, Znidaric16, Bertini21}. This is quite interesting as CFD contains density correlations which are manifestly linked with the transport.  Therefore, our results suggest that CFD captures effects of quantum information spreading in the chain that occurs faster in time than the particle transport but still slower than the ballistic dependence associated with the Lieb-Robinson bound. This implies that changes of $C_d(t)$ with respect to $c_d(t)$ that contribute to CFD are not due to a direct transfer of particle density $n_d$\footnote{Particle number $n_i$ is related to $z$ component of the spin via Jordan-Wigner transformation $n_i = S^z_i+1/2$ in a fully local fashion.} from the cut but rather are due to changes in details of dynamics of particles at distance $d$ from the cut. Such local changes of dynamics are associated with accumulated effect of interactions of many particles between the cut and the considered site. Note also that the power of $t$ versus $d$ dependence suddenly grows for large $d$. This is an artificial finite size effect resulting from the boundary conditions.

\subsection{Anderson localization}
\label{sec:AL}

\begin{figure}
    \centering
    \includegraphics[width=.85\columnwidth]{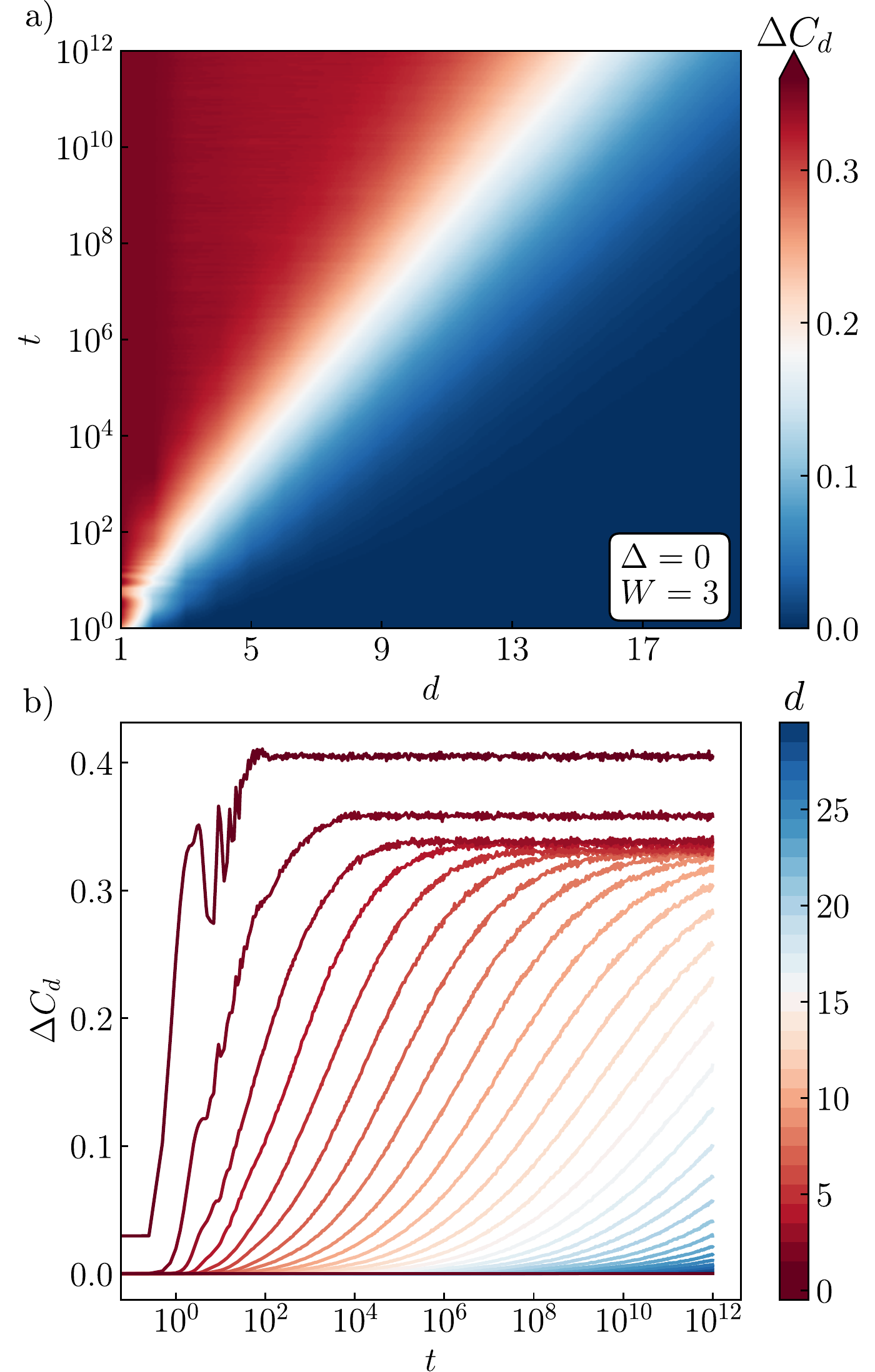}
    \caption{a) "Information log-lightcone" in the Anderson localized chain with disorder strength $W=3$. Notice the logarithmic scale for time. b) Time evolution of $\Delta C$ at different distances from the perturbed site, $L=100$, $\ell=50$. For a single site, a logarithmic growth in time can be observed until saturation. }
    \label{fig:Anderson_lightcone}
\end{figure}

We now switch off the interactions by setting ${\Delta=0}$ in \eqref{eq:xxz}. The model is Anderson-localized for any $W>0$, with localization length given by the approximate formula \cite{Potter15}
\begin{equation}
		\xiloc = \frac{3}{\log \left( 1 + W^2/4\right)},
		\label{eq:xi}
\end{equation}
which reproduces the well-known formulas for the weak-disorder limit ${\xiloc=12/W^2}$ as well as for large disorder ${\xiloc =3 / \log (W^2/4)}$.

Given the exponential localization of single particle eigenstates of the system, we expect that the spreading of quantum correlations from the center of the perturbation (i.e. the bond where the system is cut) should be exponentially suppressed compared to the previously analyzed case. Indeed, this is directly reflected in the behavior of CFD, as shown in Fig.~\ref{fig:Anderson_lightcone}a) where the "information lightcone", defined by a condition $\Delta C_d(t)=\mathrm{const}$, is visible but on the logarithmic timescale. To perform the time evolution, we map the system to a free fermion model \cite{Peschel09, Sierant22c} which, upon exact diagonalization of the resulting Hamiltonian, allows us to access the correlation function at arbitrary time $t$. Disorder averages are taken over $50000$ realizations throughout this section. Similarly, in Figure~\ref{fig:Anderson_lightcone}b) we observe that the growth of $\Delta C_d(t)$ begins at the time $\log t \propto d$. At later times, CFD increases logarithmically in time, $\Delta C_d(t) \sim \alpha \log t$, with $\alpha$ slightly decreasing with $d$, and then saturates to a certain long-time value. For $d=0,1$ this value is slightly larger than for $d\geq 2$ due to the boundary conditions (OBC) of the chain, that strongly influence the value of $c_d(t)$ for $d \approx 1$ and do not affect  $C_d(t)$. Importantly, for the Anderson insulator, in contrast to the interacting case $\Delta = 1$ discussed in the preceding section, we do not observe any decay of CFD at late times. This confirms the prediction that the latter behavior occurs only in the presence of interactions in the system. 

Phenomenology of the CFD behavior in the Anderson localized system, can be explained by the following perturbative arguments which rely on the fact of exponential localization of single particle eigenstates with localization length $\xiloc$ \eqref{eq:xi}.

\begin{figure}
	\centering
	\includegraphics[width=.85\columnwidth]{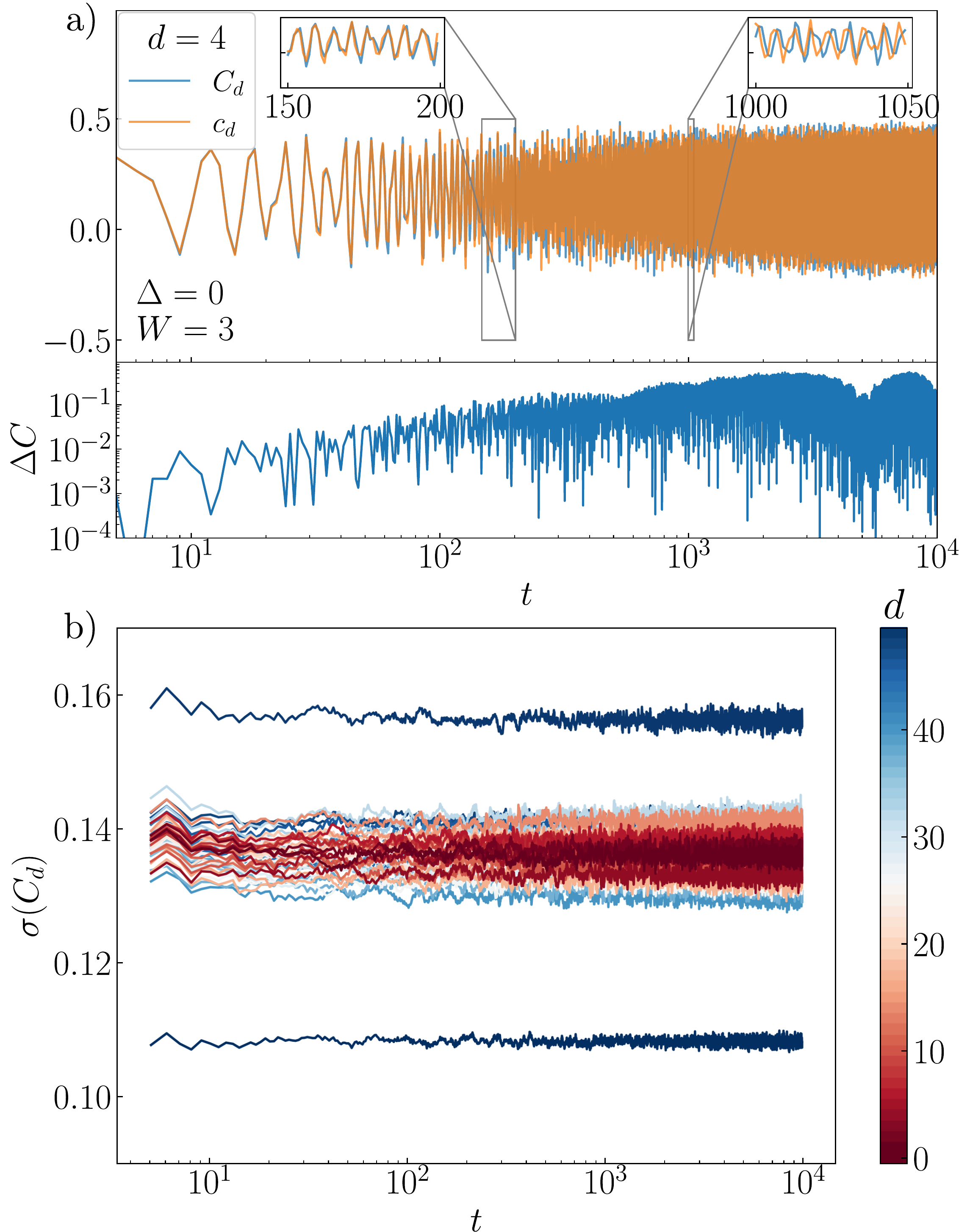}
	\caption{Single realization, panel a), of time dynamics in the Anderson model for disorder amplitude $W=3$, $L=100$, $\ell=50$. Upper panel shows both $C_d$ and $c_d$ for $d=4$. The left inset presents early time interval $t \in [150,200]$, in which $C_d$ and $c_d$ are in-phase. In the right inset, for $t\in[1000,1050]$, we observe that the dephasing between the oscillations of $C_d(t)$ and $c_d(t)$ has already occured, {yielding the major contribution to CFD, $\Delta C_d(t)$}. The panel b) shows the disorder averaged  rolling standard deviation $\sigma(C_d)$ of $C_d$. It clearly saturates fast at $d$ dependent value showing that the oscillations persists in time with a similar amplitude. 
	}
	\label{fig:Anderson_single_realization}
\end{figure}

To simplify the following considerations, let us switch to the fermionic language in which $t_{i,i+1} \equiv J\left(S^x_{i} S^x_{i+1} + S^y_{i} S^y_{i+1}\right)$ in \eqref{eq:xxz} becomes a tunneling term and the on-site disorder term is $h_i S^z_i = h_i (n_i-1/2)$, where $n_i$ is a fermionic number operator at site $i$. Furthermore, let us consider the strong disorder limit $W \gg 1$, so that the localization length is $\xiloc \ll 1$, according to \eqref{eq:xi}. In such a situation, a fermion initially localized at site $j$ at distance $d \gg \xiloc$ from the cut in the system (recall Fig.~\ref{fig:scheme}), will be oscillating with frequency $\omega_0 = J^2/(\epsilon_j - \epsilon_{j\pm1})$ between the site $j$ and a neighboring unoccupied site $j\pm1$, where $\epsilon_j \approx h_j$ in the strong disorder limit. Hence, the single site density correlation function, $c_d(t) \sim \expval{n_i(t)}$, is oscillating with frequency $\omega_0$. When we perturb the system by introducing the link between the sites $\ell$ and $\ell+1$, the frequency of oscillations of the function $C_d(t)$ changes to $\omega'_0 = J^2/(\epsilon'_j - \epsilon'_{j\pm1})$ since the energies of the states localized at sites $j$  and $j \pm 1$ are changed to $\epsilon'_j$ and  $\epsilon'_{j\pm1}$. In the strong disorder limit, the energy shifts $\Delta \epsilon_j = \epsilon'_j-\epsilon_j$ are given by the second order perturbative expressions
\begin{equation}
    \Delta \epsilon_j =\sum_{k\ne j} \frac{ |\bra{\phi_j} t_{\ell,\ell+1}   \ket{ \phi_k }|^2 }{\epsilon_j -\epsilon_k},
    \label{eq:eneshift}
\end{equation}
where $\ket{\phi_i}$ denotes an eigenstate of $H_{\ell}$ localized at site $i$. The sum over $k$ extends over the whole lattice. However, due to the exponential localization of wave functions, the matrix element in the numerator is proportional to $e^{-( |\ell-j|+|\ell-k| )/\xiloc}$. Hence, for a fixed distance from the cut $d =|\ell-j|$, the energy shift is exponentially small in $d$, $ \Delta \epsilon_j \sim e^{-d/\xiloc}$ (a shift of similar order is experienced at the site $j\pm1$). Moreover, $\Delta \epsilon_j$'s fluctuate between disorder realizations, and, in some instances the sum in \eqref{eq:eneshift} may be dominated by a single resonance term for which $\epsilon_j -\epsilon_k \ll 1$. The difference in the frequencies $\omega_0$ and $\omega'_0$ leads to a dephasing of the oscillations of $C_d(t)$ and $c_d(t)$ which takes place at time $t_{\text{deph}} \approx |\omega'_0-\omega_0|^{-1} \approx e^{d/\xiloc}$.

A comparison of $C^{\mathrm{s}}_d(t)$ and $c^{\mathrm{s}}_d(t)$ for a single disorder realization presented in Fig.~\ref{fig:Anderson_single_realization}a) immediately confirms our predictions. We observe a gradual dephasing of the oscillations of $C^{\mathrm{s}}_d(t)$ with respect to $c^{\mathrm{s}}_d(t)$ which at time $t_{\text{deph}} \approx 10^3$ leads to a saturation of CFD at a value proportional to the rolling standard deviation $\sigma(C_d)$, which, in contrast to the interacting case, saturates quickly in time, as shown in Fig.~\ref{fig:Anderson_single_realization}b). The fact that the saturation value of CFD becomes independent of $d$ for $d \gtrapprox \xiloc$ can be easily understood. When, at a given distance $d$ from the cut, the time $t$ exceeds the dephasing time $t_{\text{deph}}$, the value of CFD, $\Delta C_d$, is determined by the properties of local oscillations of fermions and properties of eigenstates in the vicinity of the site $\ell$. Both contributions become independent of $d$ upon the disorder average. Finally, the fact that $\Delta C_d(t) \sim \alpha(d) \log t$, with $\alpha$ decreasing with $d$ can be expected from the fact that the distribution of $t_{\text{deph}}$ is broad and resonances $\epsilon_j -\epsilon_k \ll 1$ in \eqref{eq:eneshift}  yield, for certain disorder realizations, growth of $\Delta C_d(t) $ even at times $ t \ll e^{d/\xiloc}$.

\begin{figure}
	\centering
	\includegraphics[width=.85\columnwidth]{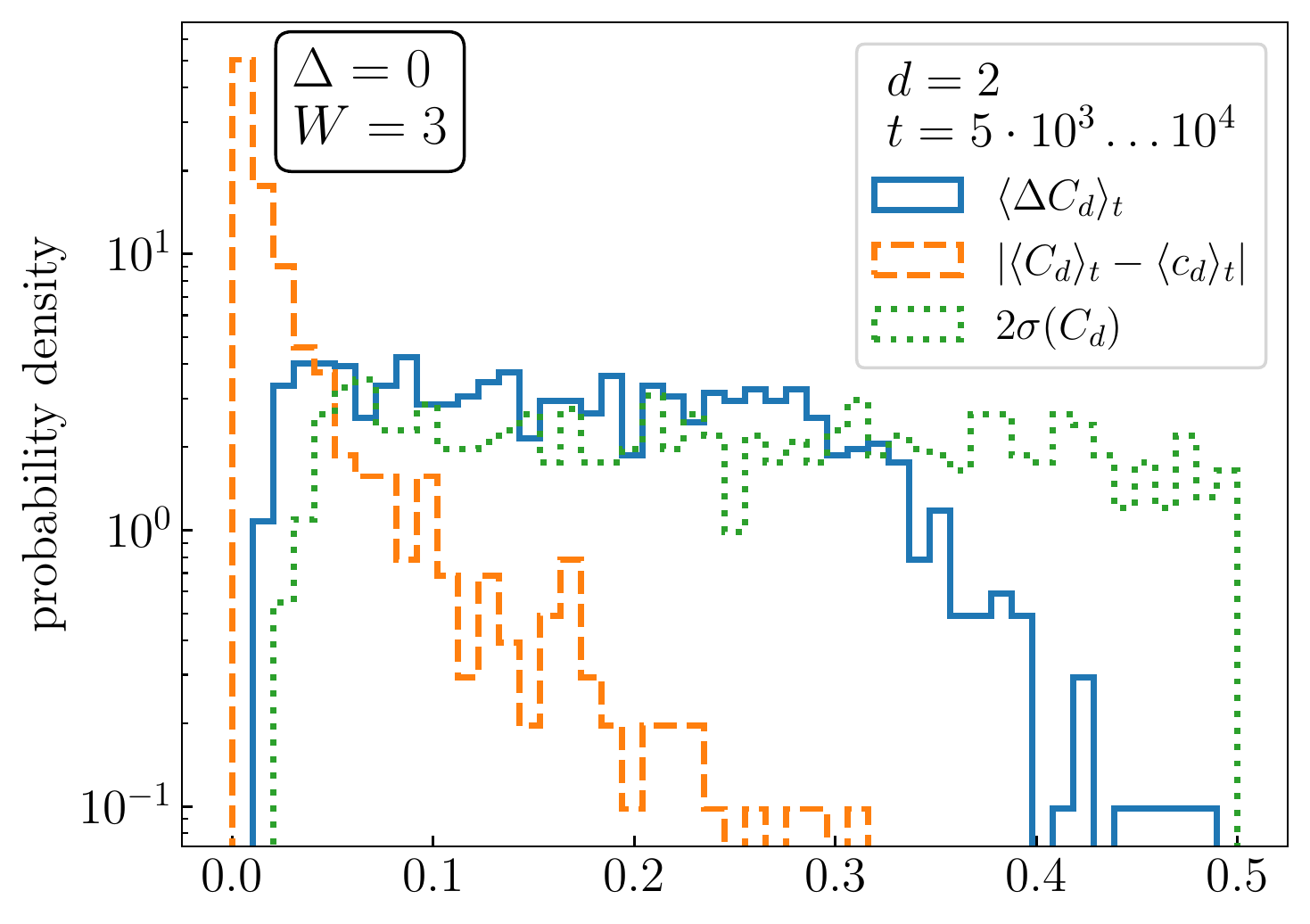}
	\caption{Histograms showing that at late times $\Delta C_d$ comes from dephasing of $C_d(t)$ versus $c_d(t)$. The system is non-interacting ($\Delta=0$), the disorder strength is set to $W=3$, and $L=16$, $\ell=14$. We compare: i) the distribution (over disorder realizations) of the absolute value difference $| \expval{C_d}_t - \expval{c_d}_t|$ between the rolling mean values of $C^{\mathrm{s}}_d(t)$ and $c^{\mathrm{s}}_d(t)$; ii) the rolling mean value $\expval{\Delta C_d}_t$ of CFD; iii) the rolling standard deviation $\sigma(C_d)$ of $C_d(t)$. Rolling window corresponds to $\Delta t=10$ tunneling times.}
	\label{fig:Anderson_histogram}
\end{figure}

Our predictions are further confirmed by the histograms in 
Fig.~\ref{fig:Anderson_histogram} where we investigate various contributions to the CFD. Firstly, we consider a difference of the rolling mean values of $C^{\mathrm{s}}_d(t)$ and $c^{\mathrm{s}}_d(t)$, take its absolute value and consider its distribution over disorder realizations and intervals of time between $t=5\cdot 10^3-10^4$. We observe that the histogram is strongly peaked at $| \expval{C}_t - \expval{c}_t| \approx \lesssim 0.03$, confirming that the difference in the average values of $C^{\mathrm{s}}_d(t)$ and $c^{\mathrm{s}}_d(t)$ has a minor contribution to the CFD $\Delta C_d(t)$. In contrast, the rolling mean value $\expval{\Delta C}_t$ of CFD has a much broader distribution that extends over to $\approx 0.4$.
The extent of this distribution is comparable to the extent of the distribution of $\sigma(C_d)$, i.e. the rolling standard deviation of the oscillations of $C_d(t)$ around the mean. This shows that the main contribution to the CFD, $\Delta C_d(t)$, is due to the dephasing mechanism rather than due to the much smaller difference of the mean values.

\begin{figure}
    \centering
    \includegraphics[width=.85\columnwidth]{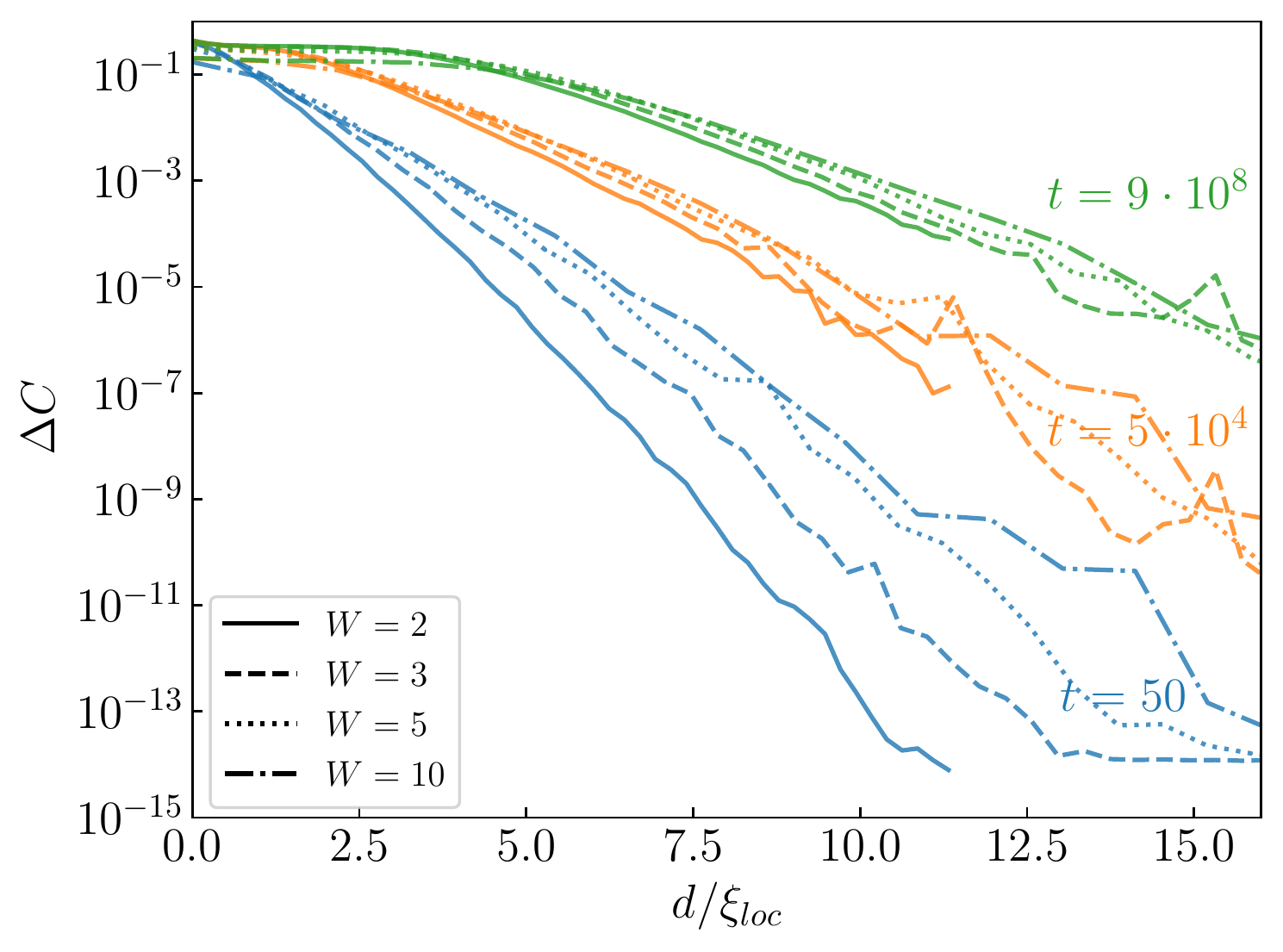}
    \caption{Exponential decrease of the CFD $\Delta C_d(t)$ with distance $d$ from the cut in the system with $L=100$, $\ell=50$, measured in units of characteristic length scale for the problem, the localization length $\xiloc$. With increasing time, universal behavior becomes more prominent. Saturation at large values of $d/\xiloc$ is due to accumulation of numerical errors in the time propagation.}
    \label{fig:Anderson_DeltaC_d}
\end{figure}

\begin{figure}
	\centering
	\includegraphics[width=.85\columnwidth]{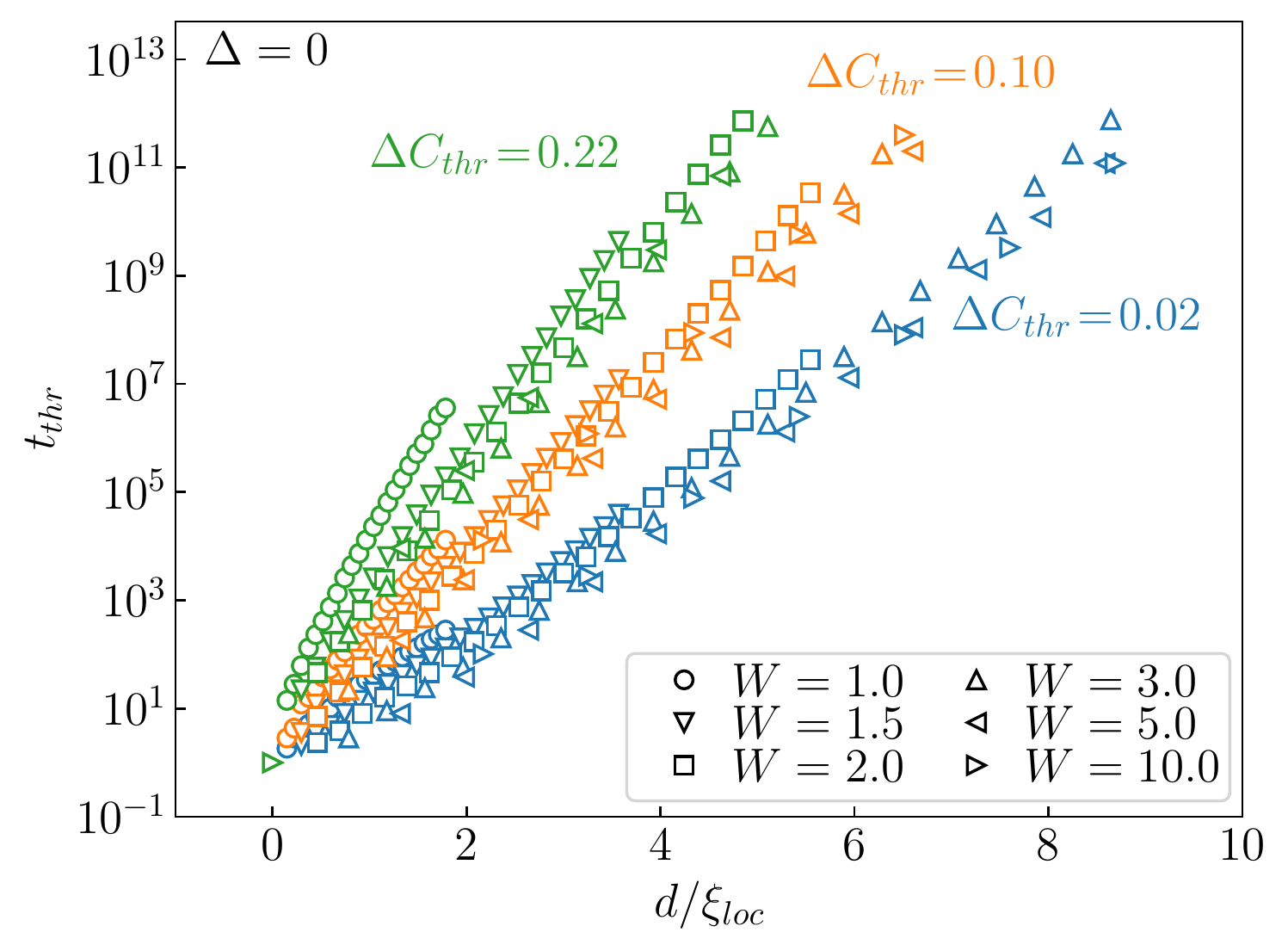}
	\caption{Time $t_{\text{thr}}$ of reaching a given value of the CFD, $\Delta C_d(t_{\text{thr}}) = \Delta C_{\text{thr}}$, in the Anderson model (Eq.~\eqref{eq:xxz} with $\Delta=0$) for different  disorder strengths $W$, $L=100$, $\ell=50$.}
	\label{fig:Anderson_t_thr}
\end{figure}

Moreover, we observe that the localization length $\xiloc$ is a universal length scale that describes the system for different values of $W$. In Figure \ref{fig:Anderson_DeltaC_d}, we show that the CFD, $\Delta C_d(t)$, at fixed time $t$, decreases exponentially as a function of $d/\xiloc$, with the agreement between the curves becoming more evident at large times. Figure $\ref{fig:Anderson_t_thr}$ shows the time $t_{\text{thr}}$ of reaching a given, fixed value of the CFD, $\Delta C_d(t)$. This time scales exponentially with $d/\xiloc$ for many values of $W$, again confirming the presence of "information log-lightcone" as well
as the fact that $\xiloc$ provides the relevant length scale. The slope in Fig.~\ref{fig:Anderson_t_thr} depends on the saturation level in $\Delta C_d$, which confirms that the distribution of dephasing time $t_{\text{deph}}$ has non-vanishing weight even at $t_{\text{deph}} \ll  e^{d/\xiloc}$. This trend, according to expectations, is more pronounced at smaller disorder strengths, e.g. $W=1$. 

Analysis performed for the Anderson model shows that in the non-interacting localized system the CFD, $\Delta C_d(t)$, changes on timescales that grow exponentially with the distance, with a universal length scale set by the localization length $\xiloc$. Due to the simplicity of the non-interacting case, we were able to pin-point the dephasing mechanism that is responsible for the main contribution to CFD. In the next section, we perform an analogous study but in presence of interactions for the MBL phase.

\subsection{MBL phase}

\begin{figure}
    \centering
    \includegraphics[width=.84\columnwidth]{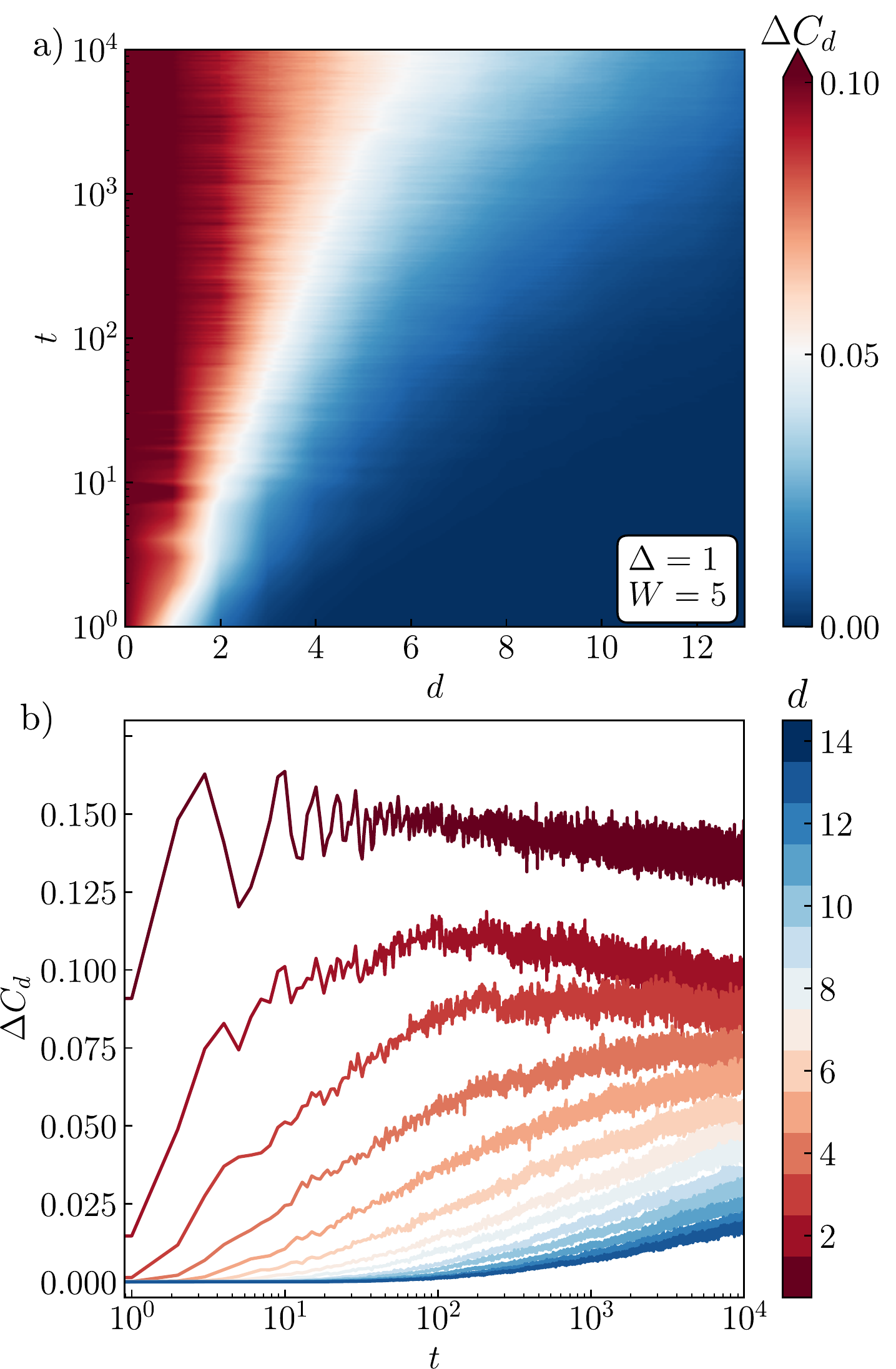}
       \caption{a) "Information log-lightcone" in the many-body localized chain with disorder strength $W=5$, $L=16$, $\ell=14$. Notice the logarithmic scale for time. b) Time evolution of $\Delta C_d$ at different distances from the cut at $\ell$. Change of monotonicity is observed due to decrease of the $C_d(t)$ oscillation amplitude with time, for details see FIG. \ref{fig:MBL_Sigmac}.}
    \label{fig:MBL_lightcone}
\end{figure}
Let us now consider the interacting case, setting $\Delta=1$ in \eqref{eq:xxz}. We start by condsidering $W=5$, the disorder amplitude which is well within the MBL regime for small system sizes analyzed here (recall that $W_c\approx 3.5$ is the estimation \cite{Luitz15} of the position of ergodic-MBL crossover at $L\approx 20$). To calculate the time evolution we use the Chebyshev expansion of the time evolution operator \cite{Fehske08}.
Similarly to the non-interacting case, we observe a logarithmic lightcone for the CFD $\Delta C_d(t)$ for $\Delta = 1$, c.f. Fig.~\ref{fig:MBL_lightcone}a). Another feature shared by the interacting and the Anderson localized   
models is the linear growth of $\Delta C_d(t)$ with $\log t$ presented in Fig.~\ref{fig:MBL_lightcone}b). There is, however, a notable difference in the long-time behavior of $\Delta C_d(t)$ between the two cases considered. While in the AL case, the values of $\Delta C_d(t)$ always grow and saturate at a constant value given by the $C_d(t)$ long-time oscillation amplitude (except for $d=0,1$ that saturate at slightly larger values due to OBC around site $\ell$, see discussion in section \ref{sec:AL}), in the MBL case, at small values of $d$, $\Delta C_d(t)$ first grows, then saturates and finally at still longer times decreases towards an equilibrium value. The phenomenology of the initial, logarithmic in time, growth of the CFD at all distances $d$ can be explained in the same manner as for the AL case. Here, the local integrals of motion (LIOM) \cite{Serbyn13b,Huse14,Ros15, Chandran15} play the role of projectors on localized eigenstates, and the main contribution to CFD comes from the dephasing mechanism. 
 
\begin{figure}
    \centering
    \includegraphics[width=.85\columnwidth]{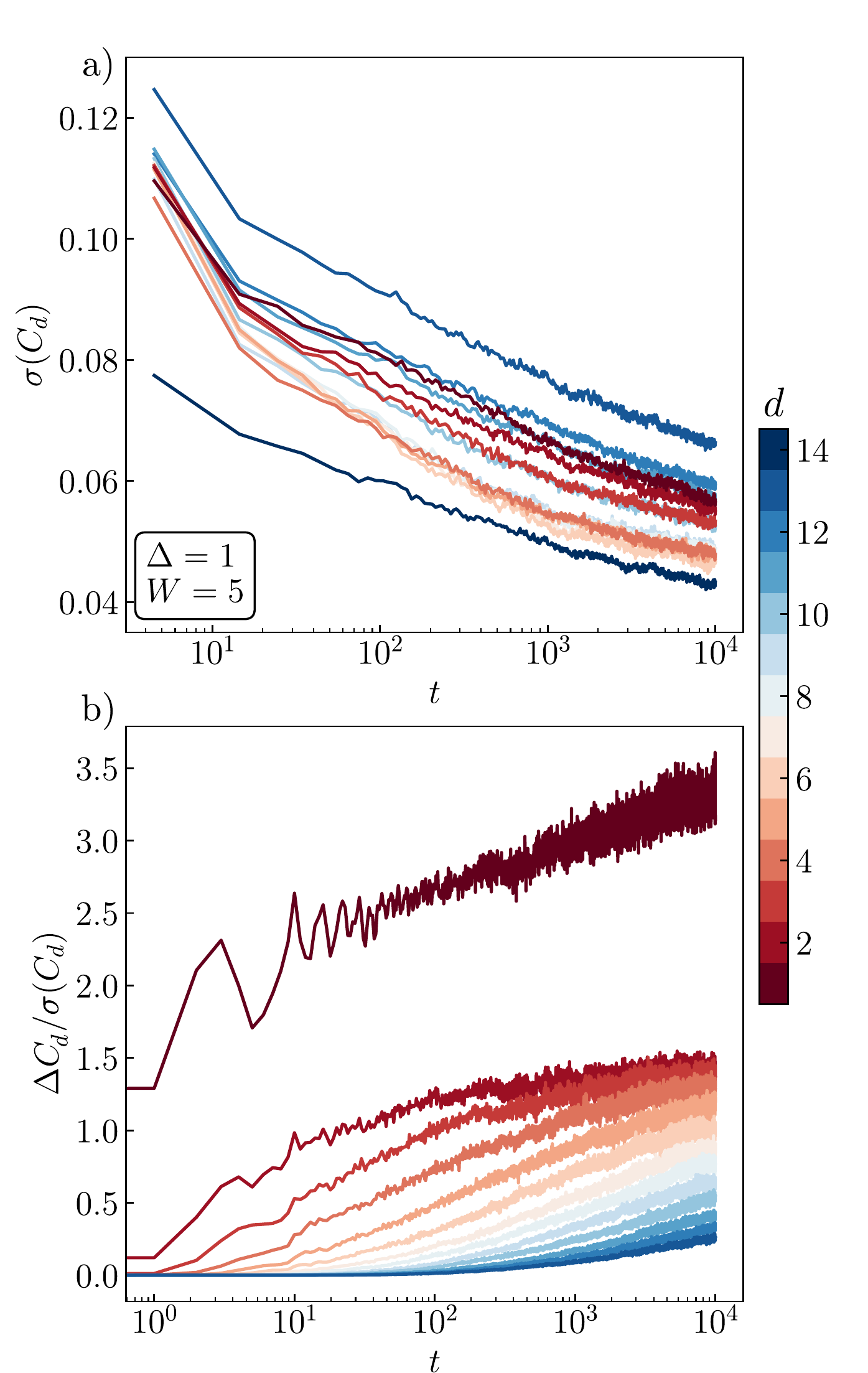}
\caption{a) Rolling standard deviation of $C_d(t)$ with time window of size $\Delta t=10$ and system size $L=16$, cut at $\ell=14$. Deviation from the mean decreases with time, explaining the decrease of $\Delta C_d$ at long times observed in  Fig.~\ref{fig:MBL_lightcone}. 
b) $\Delta C_d(t)$ in units of the rolling standard deviation. The decreasing trend is inverted.}
    \label{fig:MBL_Sigmac}
\end{figure}

The change in monotonicity of CFD at later times is analogous to the previously analyzed ergodic case, c.f. Sect.~\ref{sec:erg}. It can be technically explained by an inspection of the oscillation amplitudes of $C_d(t)$ and $c_d(t)$ calculated as the rolling standard deviations of the signal $\sigma(C_d(t))$ for the rolling window of $\Delta t=10$, according to Eq.~\eqref{eq:roll}. In contrast to the Anderson localized phase, where $\sigma(C_d(t))$ strictly approaches a constant as a function of time $t$ (see the bottom panel in Fig.~\ref{fig:Anderson_single_realization} for this effect), temporal fluctuations of $C_d(t)$ decrease logarithmically in time in the interacting model. This decay is significant, as $\sigma(C_d(t))$ decreases by a factor of even $30\%$ between short times of order $10$ and longer times of order $10^4$, see Fig.~\ref{fig:MBL_Sigmac}a). The decay of $\sigma(C_d(t))$ reflects the fact that oscillations present in $C_d(t)$ (and $c_d(t)$) actually decay in amplitude with time, as observed earlier for interacting systems in \cite{Serbyn14, Nandy21}.
To compensate for this effect we may inspect the dimensionless rescaled CFD $\Delta C_d/\sigma(C_d(t))$ shown in Fig.~\ref{fig:MBL_Sigmac}b). It grows monotonically in time and does not saturate before time $t=10^4$. This suggests that slow but non-trivial dynamics of the density in the system persist to the longest times considered.

\begin{figure}
    \centering
    \includegraphics[width=.85\columnwidth]{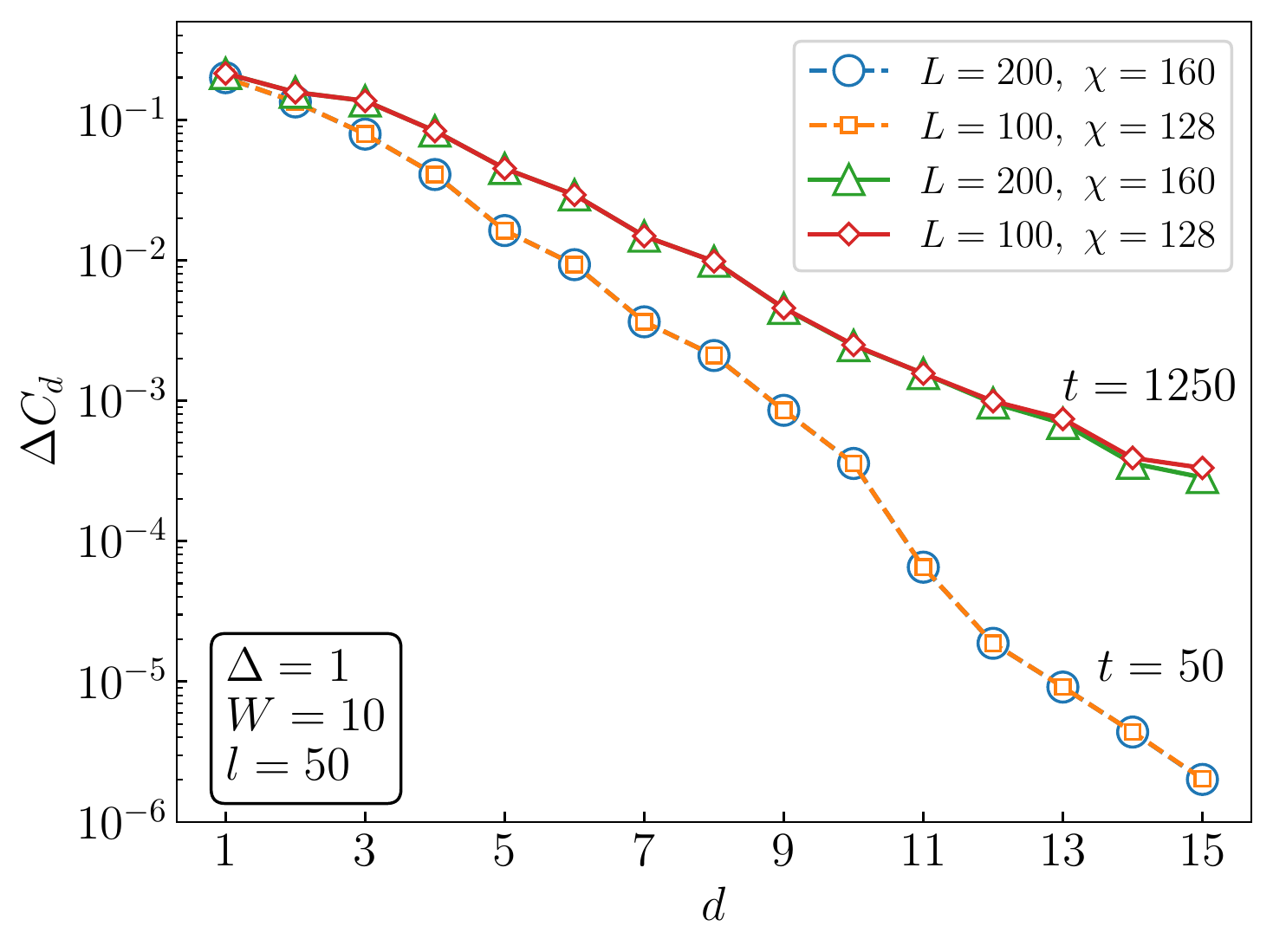}
\caption{Exponential decrease of the CFD $\Delta C_d$ with the distance from the cut $d$ for XXZ chain with $L=200,100$, with the cut at $\ell=50$. The disorder strength is fixed as $W=10$ and CFD is calculated at times $t=50$ and $t=1250$. Time evolution of the system was found using tensor network TDVP algorithm. Average is calculated over $1008$ disorder realizations.}
    \label{fig:TDVP}
\end{figure}

Finally, we consider time evolution of larger systems ($L=100,200$) calculated with the TDVP algorithm \cite{Haegeman11, Koffel12, Haegeman16, Chanda19}. We choose a relatively strong disorder ${W=10}$, which allows us to calculate the time evolution of the system to times of the order of $10^3 J^{-1}$ with  modest bond dimensions $\chi < 200$. In Fig.~\ref{fig:TDVP} we plot the CFD as function of the distance $d$ from the cut in the system, observing an exponential decay $\Delta C_d \sim e^{-d / \xi_{\text{MBL}}(t)}$. The characteristic lengthscale $\xi_{\text{MBL}}(t)$ slowly increases with time $t$, consistently with non-trivial dynamics of imbalance at those time scales reported in \cite{Sierant22c}. Thus the analysis of CFD in the MBL regime does not bring a definitive answer about the locality of the dynamics that would be signified by $\xi_{\text{MBL}}(t) \stackrel{t\to\infty }{\to} \mathrm{const}$. Nevertheless, our results emphasize the analogies in the behavior of CFD between the MBL regime and Anderson insulator, showing similar degrees of locality in dynamics of the systems at finite times.

\section{Relation to other locality measures}
The CFD, $\Delta C_d(t)$ is a local probe of dynamics and can be compared with other measures of locality, for instance, with the entanglement entropy. We note, however, that the latter quantity is much more difficult to access experimentally.

In particular, the von Neuman bipartite entanglement entropy $S_{d}(t)$ of subsystem $A$ consisting of sites $i\in[1,...,(\ell-d)]$ can be written as a sum of two terms (see e.g. \cite{Schuch04,Schuch04b,Donnelly12,Turkeshi20,Lukin19,Sierant19c,Yao21,Sierant22c} for other applications):
\begin{equation}
    S_{d}(t) = \Snum(t) + \Sconf(t),
\end{equation}
with the configurational entropy given by
\begin{equation}
    \Sconf(t) = -\sum_n p(n) \Tr\left[ \rho(n) \ln \rho(n)\right],
\end{equation}
where $p(n)$ denotes the probability that $n=\sum_{i=1}^{\ell-d}S^z_i$ and $\rho(n)$ is the block of the reduced density matrix in sector with magnetization $n$, while the number entropy reads
\begin{equation}
    \Snum(t)=-\sum_n p(n) \log p(n).
\end{equation}
In the spinless fermion description of the model, the number entropy contains information about the distribution of the number of particles in one part of the chain, with a total number of particles conserved, while the configurational entropy measures the number of possible particle arrangements within the subsystem. 

\begin{figure}
    \centering
    \includegraphics[width=1.03\columnwidth]{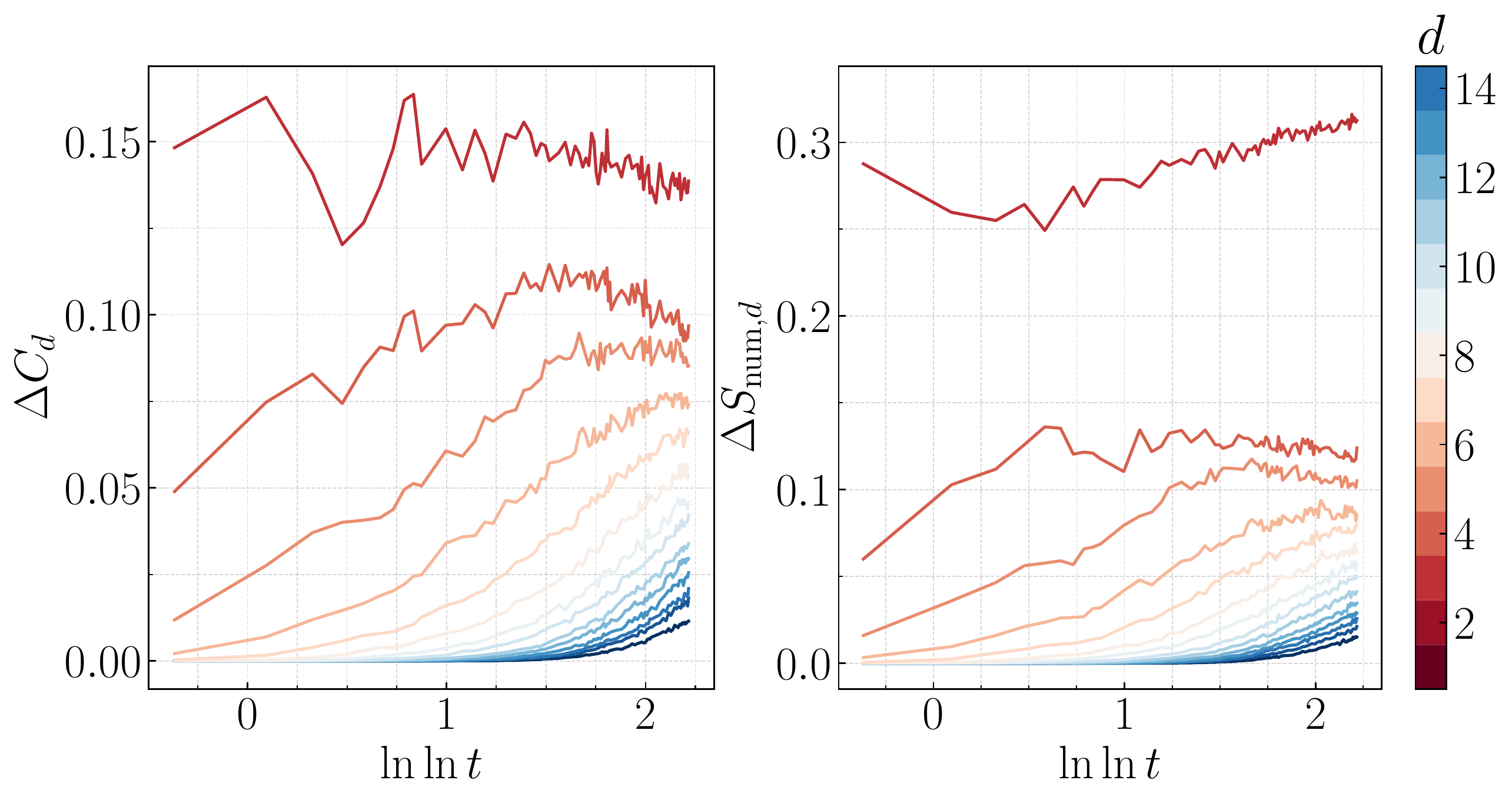}
    \caption{Side-by-side comparison of $\Delta C_d(t)$ and the difference in the number entropy $\Delta \Snum(t)$ for system size $L=16$, cut at $\ell=14$, see text.}
    \label{fig:AL_S}
\end{figure}

We investigate the absolute differences of the two entropies $\Delta \Sconf(t)$, $\Delta \Snum(t)$ after performing the cut between sites $\ell$ and $\ell+1$, in the same manner as we investigate $\Delta C_d(t)$.
In Fig.~\ref{fig:AL_S} we find that the time dependence of $\Delta C_d(t)$ is compatible with the difference of the number entropy $\Delta \Snum (t)$. This is expected $\Delta \Snum (t)$: both $C_d(t)$ and $\Snum(t)$ provide information about a flow of particles between two subsystems.
For small $d$, i.e., when the cut is close to the site studied, $\Delta C_d(t)$ and $\Delta \Snum (t)$ dynamics strongly differ, apparently the correlation between both quantities is small and affected by local dynamics in the vicinity of the cut. On the other hand, for sufficiently large $d$, say $d>6$, $\Delta C_d(t)$ time dependence mimics that of $\Delta \Snum (t)$. The latter
seems to grow slowly in time with the growth being similar to $\ln \ln t$ time dependence resembling the recent findings of such a growth in the number entropy itself \cite{Kiefer20,Kiefer21}. Since the matter seems to be controversial \cite{Ghosh22} especially due to the difficulty of treating really large system sizes, the similarity between  $\Delta C_d(t)$ and   $\Delta \Snum (t)$ time dynamics for larger $d$ suggests than one could also use the former in experimental studies to possibly resolve the issues related to number entropy.

Additionally, we do not expect that CFD may bring some useful information on the configurational entropy that provides, let us recall, information about possible particle rearrangements in one of the subsystems. This inherently quantum information may be recovered from spin-spin correlations at different sites \cite{DeChiara18, Schweigler17}, the analysis of the behavior of this quantity is beyond the scope of the present work.

\section{Conclusions}
\label{sec_con}
We have introduced CFD, a local quantity that allows to quantify the degree of locality of dynamics of quantum many-body systems. The CFD, $\Delta C_d(t)$, compares time evolution of single site density correlation functions in two situations: when the 1D system is comprised of $L$ sites and when the system is cut at site $\ell < L$. A vanishing CFD at a distance $d$ from the cut indicates that the degrees of freedom residing at sites beyond the cut do not affect the dynamics at the considered site. Conversely, a non-vanishing value of $\Delta C_d(t)$ for a certain $d$ quantifies the degree of influence of sites beyond the cut on the dynamics at the considered distance from the cut. The determination of CFD requires a single site resolution for measurement of density correlation functions and a single site control over the Hamiltonian of the system. Both conditions are satisfied by present day experiments with ultra-cold systems.

In order to benchmark the behavior of CFD, $\Delta C_d(t)$, in various dynamical phases of matter, we have studied the time evolution of disordered XXZ spin chain. Varying the parameters of the system, we tuned the system between an ergodic phase, Anderson insulator phase and MBL regime. We have shown that the dynamics of the considered dynamical regimes are characterized by distinct behaviors of CFD. In accordance with the expectation that the dynamics quickly gets non-local in the ergodic phase, we have observed a quick increase of CFD, finding that the condition $\Delta C_d(t)=$const implies a power law relation between the distance from the cut, $d$, and time, $t$, at which CFD reaches a given value. For the Anderson localized system, we have found that CFD increases logarithmically slowly in time. Consequently, the condition ${\Delta C_d(t)=\mathrm{const}}$ yields the relation $t\sim e^{d/\xiloc}$ between the time and space, where $\xiloc$ is the localization length of single particle eigenstates that becomes the relevant lengthscale for the problem. Finally, we have demonstrated a similar dependence for MBL regime. In the latter case, we observed a slow increase in time of the characteristic length scale that governs the exponential decay of CFD, in analogy with earlier studies of dynamics in MBL regime.

Our study demonstrates the utility of CFD in quantifying the degree of locality of quantum dynamics of many-body systems. Observing the diffusive behavior in the ergodic phase and exponentially slow dephasing in the localized case, we have shown that CFD is sensitive to dynamics of the density of particles, but remains oblivious to more complex quantum correlations in the system.  Natural possible directions for future work include: extending CFD to systems of bosons, understanding its behavior in the absence of symmetries such as the conservation of the total number of particles and studying analogous differences of other, possibly nonlocal, observables such as two-point correlation functions.
Another avenue for future research is to investigate the behavior of CFD in disordered many-body systems with real space renormalization group approaches \cite{Vosk13, Monthus18, Ruggiero22}.

{\it Note} The numerical data presented in this work are freely available from  \url{https://chaos.if.uj.edu.pl/ZOA/opendata/} or from the authors upon a reasonable request.

\vfill\null

\acknowledgements    
We thank F. Evers for the illuminating discussions.
The numerical computations have been possible thanks to PL-Grid Infrastructure. The works of T.S. and J.Z. have been realized within the Opus grant 2019/35/B/ST2/00034, financed by National Science Centre (Poland). P.S. and M.L. acknowledge support from: ERC AdG NOQIA; Ministerio de Ciencia y Innovation Agencia Estatal de Investigaciones (PGC2018-097027-B-I00/10.13039/501100011033, CEX2019-000910-S/10.13039/501100011033, Plan National FIDEUA PID2019-106901GB-I00, FPI, QUANTERA MAQS PCI2019-111828-2, QUANTERA DYNAMITE PCI2022-132919, Proyectos de I+D+I “Retos Colaboraci\'on” QUSPIN RTC2019-007196-7); MCIN Recovery, Transformation and Resilience Plan with funding from European Union NextGenerationEU (PRTR C17.I1); Fundaci\'o Cellex; Fundaci\'o Mir-Puig; Generalitat de Catalunya (European Social Fund FEDER and CERCA program (AGAUR Grant No. 2017 SGR 134, QuantumCAT \ U16-011424, co-funded by ERDF Operational Program of Catalonia 2014-2020); Barcelona Supercomputing Center MareNostrum (FI-2022-1-0042); EU Horizon 2020 FET-OPEN OPTOlogic (Grant No 899794); National Science Centre, Poland (Symfonia Grant No. 2016/20/W/ST4/00314); European Union’s Horizon 2020 research and innovation programme under the Marie-Skłodowska-Curie grant agreement No 101029393 (STREDCH) and No 847648 (“La Caixa” Junior Leaders fellowships ID100010434: LCF/BQ/PI19/11690013, LCF/BQ/PI20/11760031, LCF/BQ/PR20/11770012, LCF/BQ/PR21/11840013).

%\bibliography{ref_02_20}

\end{document}